# Characteristic market behaviors caused by intervention in a foreign exchange market

Takayuki Mizuno[1], Yukiko Umeno Saito[2], Tsutomu Watanabe[2,3], and Hideki Takayasu[4]

[1] Department of Computational Intelligence and Systems Science, Interdisciplinary Graduate School of Science and Engineering, Tokyo Institute of Technology, 4259 Nagatsuta-cho, Midori-ku, Yokohama 226-8502, Japan
[2] Fujitsu Research Institute Inc., 1-16-1 Kaigan, Minato-ku, Tokyo 105-0022, Japan
[3] Institute of Economic Research, Hitotsubashi University, 2-1 Naka, Kunitachi City, Tokyo 186-8603, Japan
[4] Sony Computer Science Laboratories, 3-14-13 Higashi-Gotanda, Shinagawa-ku, Tokyo 141-0022, Japan

**Summary.** In foreign exchange markets monotonic rate changes can be observed in time scale of order of an hour on the days that governmental interventions took place. We estimate the starting time of an intervention using this characteristic behavior of the exchange rates. We find that big amount of interventions can shift the averaged rate about 1 yen per 1 dollar in an hour, and the rate change distribution becomes asymmetric for a few hours.

## 1. Introduction

The central banks intervene in foreign exchange markets in order to stabilize the currency exchange rates. The amount of one transaction is typically several million dollars in the yen-dollar market, while the amount of intervention by the bank of Japan sometimes exceeds 10 billion dollars in one day. Due to such extraordinary big amount the intervention is generally expected to show remarkable influence on the market. For the basic study of foreign exchange markets it is important to clarify the influence of interventions quantitatively and statistically.

Among various studies on foreign exchange markets the effects of interventions have been investigated by using the daily data [1][2]. In order to clarify the response of markets to an intervention in more detail, we investigate average rate shifts after the intervention in the time scale of several minutes.

As the intervention starting times of the Bank of Japan are not announced, we first estimate the starting time using characteristics of rate changes on intervention days. By analyzing the yen-dollar market's tick data for 10 years comparing the information of intervention about amounts and dates, we find statistical laws of rate changes for typical intervention influences.

## 2. Asymmetrical rate changes on intervention days

In Fig.1 we show yen-dollar fluctuations on 2/11/1994 as a typical example of an intervention day. It is announced that the United States' governmental bank intervened in the foreign exchange market from 11:00AM on this day. Right after the start of the intervention the exchange rate went up nearly monotonically for about 30 minutes. We can find similar rate changes also on other intervention days. The monotonic rate changes cause an asymmetric probability density distribution of rate changes on the intervention day.

We investigate the skewness of rate changes on intervention days. Denoting yen-dollar rate at a time $t$ by $P(t)$ and its change with time interval $T$ by $dP(t,T) = P(t+T) - P(t)$ the skewness is defined by the following formula:

$$\text{Skewness} = \frac{\left\langle \left( dP(t,T) - \langle dP(t,T) \rangle \right)^3 \right\rangle}{\sigma^3}, \qquad (1)$$

where $\sigma$ is the standard deviation of the rate change. We plot the skewness as a function of time scale $T$ in Fig.2. White diamonds ( ) denote the averaged skewness on non-intervention days. The rate change distributions on the non-intervention days are nearly symmetric as known from the value of the skewness being around zero [3]. Black diamonds ( ) show the averaged skewness on days of "yen-selling intervention". Within the time scales between 10 and 60 minutes, non-zero skewness is clearly observed on the intervention days. Namely, the rate change distribution on intervention days is asymmetric in this time scale.

## 3. Intervention starting time estimation

The number of Japanese interventions is much larger than that of any other country. For example, in the yen-dollar exchange market from 1991 to 2004, the total number of Japanese interventions was 342, while that of the United States' interventions was only 10. In order to analyze statistical properties we focus on Japanese interventions. The dates of Japanese interventions are now announced and each amount is also announced, however, the starting times are never announced. Therefore, we have to develop a method of estimating the starting time of Japanese intervention from given rate change data.

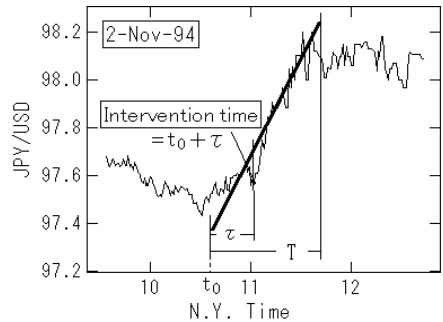

**Fig.1** An example of intervention in the yen-dollar exchange market. The figure shows the rate in 2/11/1994. America intervened at 11:00AM.

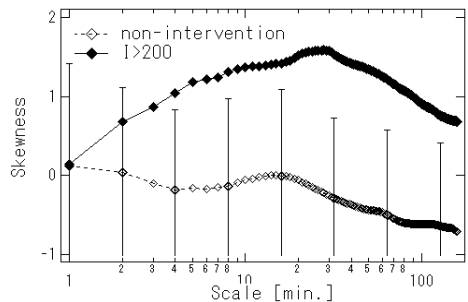

**Fig.2** Skewness of the rate change on a time scale between 1 minute and 3 hours.   is the averaged skewness by non-intervention days and   is the averaged skewness of days of "yen-selling intervention" more than 200 billion yen. Error bars are estimated by the standard deviation of the skewness for non-intervention days.

On an intervention day a large rate change can often be found in the time scale between 10 to 60 minutes as mentioned. We assume that such a large rate change occurs due to the intervention and define a characteristic time $t_0$ that gives the starting time of the largest rate change smoothed over time scale of 60 minutes on each intervention day as indicated in Fig.1. In Table.1 we show a relationship between the intervention time and the characteristic time $t_0$ through the period in which the starting times of the United States' interventions were announced. We find that the actual intervention starting time is given $t_0 + 20$ minutes. We apply this estimation method for the estimation of starting times of Japanese interventions.

**Table.1** The United States' interventions from 1994 to 2000. The table shows intervention dates, intervention times, characteristic times $t_0$, and the intervened market. The characteristic time $t_0$ is defined during a time when the largest rate change for 60 minutes occurred as shown in Fig.1.

| Date | time | $t_0$ | $t_0$+20minutes | Market |
|---|---|---|---|---|
| 29-Apr-94 | 11:30 | 11:20 | 11:40 | Yen-Dollar |
| 4-May-94 | 8:30 | 8:20 | 8:40 | Yen-Dollar |
| 24-Jun-94 | 9:30 | 9:20 | 9:40 | Yen-Dollar |
| 2-Nov-94 | 11:00 | 10:40 | 11:00 | Yen-Dollar |
| 3-Nov-94 | 11:00 | 10:40 | 11:00 | Yen-Dollar |
| 3-Mar-95 | 9:10 | 8:30 | 8:50 | Yen-Dollar |
| 3-Apr-95 | 11:20 | 13:50 | 14:10 | Yen-Dollar |
| 5-Apr-95 | 10:20 | 10:00 | 10:20 | Yen-Dollar |
| 17-Jun-98 | 7:55 | 7:35 | 7:55 | Yen-Dollar |
| 22-Sep-00 | 7:11 | 7:10 | 7:30 | Dollar-Euro |

## 4. Influence of intervention

We introduce a conditional average: $\langle P(t_0+dt)-P(t_0) | I_{c1} < I < I_{c2} \rangle$, that is the average of rate changes $P(t_0(n)+dt)-P(t_0(n))$ for yen-selling intervention of which amount is in the range of $I_{c1}<I<I_{c2}$. Here, $t_0(n)$ denotes the characteristic time of the $n$-th intervention. We show the results $\langle P(t_0+dt)-P(t_0) | I_{c1} < I < I_{c2} \rangle$ for three ranges of the intervention amounts in Fig.3. The time $dt = 20$ minutes is expected to give the starting of interventions. For big size interventions of $I > 200$ billion yen the average rate shifts monotonically for about an hour and it becomes nearly flat after that. From Fig.3 we find that the average rate shift depends on the amount of intervention. Although the shift is within the error bars for $I < 200$ billion yen, the over all rate shift due to the intervention is approximated by a linear function,

$$\langle P(t_0+dt)-P(t_0) | I \rangle \simeq I \times 10^{-3}, \qquad (2)$$

where, $dt = 190$ minutes and a unit of the intervention price $I$ is one billion yen.

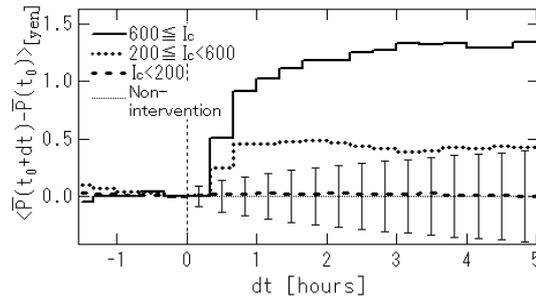

**Fig.3** Average rate change after intervention times. The top (solid) line indicates cases in which the interventions are larger than 600 billion yen, the middle line represents cases of interventions from 200 billion yen up to 600 billion yen, and the bottom line indicates cases in which interventions are less than 200 billion yen. Error bars are standard deviations of a rate change on non-intervention days.

## 5. Discussion

We introduced an estimation method of intervention times and clarified influences of the intervention on a foreign exchange market. The exchange rate drifts for about an hour after the start of an intervention time. After an hour the rates fluctuate randomly with no drift. The shift of the averaged rate is approximated by a linear function of the intervention price. For example, typically an intervention of size 1 trillion yen can shift 1 yen in the yen-dollar exchange rates.

## Acknowledgement

The present authors would like to show appreciation to Hiroyuki Moriya in Oxford Financial Education for providing us with the CQG data of high-frequency exchange rate, Prof. Tohru Nakano for stimulating discussions. T. Mizuno is supported by Research Assistant Fellowship of Chuo University, and the Ministry of Education, Science, Sports and Culture, Grant-in-Aid for JSPS Fellows.

## References

[1] T. Ito, Is Foreing exchange intervention effective? The Japanese experiences in the 1990s, NBER Working Paper No. 8914, 2002.
[2] A. P. Chaboud and O. F. Humpage, An analysis of Japanese foreign exchange interventions, 1991-2002, Federal Reserve bank of Cleveland, Working Paper 03-09, 2003.
[3] T. Mizuno, S. Kurihara, M. Takayasu, H. Takayasu, Analysis of high-resolution foreign exchange data of USD-JPY for 13 years, Physica A 324, 296-302, 2003.